\documentstyle[amsfonts,preprint,prl,aps,epsfig]{revtex}

\newcommand{\beq}{\begin{equation}}
\newcommand{\eeq}{\end{equation}}

\newcommand{\ket}[1]{| #1 \rangle}
\newcommand{\bra}[1]{\langle #1 |}
\newcommand{\proj}[1]{\ket{#1}\! \bra{#1}}
\newcommand{\inner}[2]{ \langle #1 | #2 \rangle}

\tightenlines

\begin{document}
%\draft 
\title{Reversible Mapping for Tree Structured Quantum Computation}
 
\date{November 30, 1997}
\author{Wojciech Burkot}
\address{Institute of Automatics,Electrical Engineering Dept.\\
Technical University AGH, Mickiewicza 30, PL30-059 Krak\'ow, Poland\\
email: burkot@ia.agh.edu.pl\\
}
\maketitle
 
\begin{abstract}
A  hierarchical, reversible mapping between levels of tree structured 
computation, applicable for structuring the Quantum Computation algorithm 
for ${\cal NP}$-complete  problem is presented.
It is proven  that confining the state of a quantum computer to a subspace
of the available Hilbert space, where states are consistent with the problem
constraints, can be done in polynomial time.
The proposed mapping, together with the method of the state reduction
can be potentially used for solving  ${\cal NP}$-complete problems in
polynomial time.
\end{abstract}
\narrowtext

\paragraph*{Motivation}
Expected running time scaling polynomially with input size, displayed
by Shor's factoring  algorithm \cite{Shor} and, to even greater extent, 
the more recent results \cite{QNoQC,Bra} give evidence  that Quantum Mechanics 
offers a computation model more powerful  than a classical one,
yet still physically feasible.

An application this power for simulating  nondeterminism associated with
${\cal NP}$-complete problems  had been  discouraged by Bennett et al.  in  \cite{Ben}.
However, 
Theorem 2 of  \cite{Ben}, which  can be interpreted as a ``no-go''
proof for unstructured  attempts to solve
${\cal NP}$-complete problems  in polynomial time,
assumes particular quantum computation consisting of unitary
evolution of simple input state followed by a single measurement of a state
of a system.

We will follow different approach, patterned after error correction, with 
repeated measurements forcing  projections of a system state onto a 
subspace spanned on eigenvectors of particular eigenvalue
of the Hermitian  operator representing measurement.

In  remaining part of this paper, after choosing a known ${\cal NP}$-complete
problem as our object of study,  we will present the deterministic, 
data independent mapping  between the levels of a solution construction tree,
thus providing a method for structuring quantum algorithms for 
${\cal NP}$-complete problems.  

\paragraph*{Sample Problem}
Let us consider a ${\cal NP}$-complete 
combinatorial optimization problem known as a symmetric 
Traveling Salesman Problem (TSP)  defined as follows:
\newline
Given a weighted complete graph $G=(V,E,w)$,where $V$ is a set of 
vertices $v_k,\; k=1.. n$,
E is a set of edges, \mbox{$\forall v_i,v_k \in V\; \{v_i,v_k\}\in E$} 
and a function $w: E\rightarrow {\cal Z} $ is symmetric:
 \mbox{$w_{i,k}=w(e_{i,k})\equiv w(e_{k,i}),$}
\newline
find a Hamiltonian cycle $t$ 
(a closed tour through the graph, visiting each vertex exactly once) with 
the minimal total weight:
\begin{equation}
    W(t)=  w(\{v_{t(n)},v_{t(1)}\}) + \sum_{k=1}^{n-1} w(\{v_{t(k)},v_{t(k+1)}\}) 
\end{equation}  
%Without loss of generality we can assume all $w_{i,k}$ positive.  

Instead of coding each  cycle as a list of vertices,
we will code it as a set of its edges, since in this case ordering information
is coded locally, reducing number of bits modified at a cycle modification.
We can represent each edge present in a cycle by a single set bit in a list  of
all $n(n-1)/2$ edges. The ordering of the list is such that the edges are 
grouped by the higher vertex number present in the edge and,
within the group, by the number of the other  vertex of the edge: 
\mbox{$({e_{2,1},e_{3,1},e_{3,2},e_{4,1},\dots })$ }.
In such a way, incomplete potential solutions i.e. cycles over first $m$
vertices of a $n$-vertex  problem have all bits above $m(m+1)/2$ reset to 0. 
We  need ${\cal O}(n^2)$ bits to code a $n$-vertex cycle,
as compared with   ${\cal O}(n\log(n))$ of more common coding as 
a list of vertices.
Such a redundancy can be used for rudimentary error control,
but the main motivation is  robustness, universality and  particular 
simplicity of unitary operator used for mapping between levels of a  
construction tree. 

From each cycle of $m$ vertices we can construct $m$ cycles connecting 
$m+1$ vertices  by breaking it at one of  its $m$ edges, and inserting
the vertex $m+1$ at the breaking point.
We reset the relevant bit to zero (since the edge $e_{i,j}$ we break at
no longer exists in a resulting cycle ) and  replace it  by two bits
representing edges which connect broken ends of the cycle of a previous
level to the newly inserted vertex: $e_{m+1,j},$ $e_{m+1,i}$.

\paragraph*{Reversible Mapping for TSP}

A quantum computation method of producing all the   
permutations of a given set of elements (all vertices $v_i\in V $ in a case 
of TSP) has been already provided  in \cite{Bar}, 
however, it requires resources 
scaling exponentially with number of elements in a set.

%Our construction will be based on two facts of general nature: 
%\begin{enumerate}
%\item
%In order to use a deterministic mapping between levels of 
%tree structured computation we must provide an auxillary system 
%with sufficient number of states  to compensate 
%for a difference of entropy between the involved levels. 
%\item
%The evolution of isolated quantum system is, by unitarity,  
%deterministic. The only way we can simulate nondeterminism in such a
%system is through measurement.
%\end{enumerate}

It is interesting to note that latest results in Quantum Computation 
are not presented in 
Quantum Turing Machine formalism. This, in author opinion, reflects
the growing belief that if we are ever able to build a quantum computer it
will rather be in a form of uniform array of identical, locally interacting  
elements with quantum degrees of freedom \cite{Mar,SBen}
than the counterpart of a classical computer. We will follow this approach.

We  start with  the only tour connecting three
vertices represented by a state $\ket{1_{2,1} 1_{3,1} 1_{3,2}\bbox{ 0 }}$,
where $\bbox {0}$  
is used to denote that all the remaining qbits in all remaining groups
are 0. In subsequent equations we will drop the indices for shorter  notation, 
while preserving the same convention for the order of qbits.

Let us denote $l-$th bit of a pure state $|s>$ by $s\{l\}.$ 
Clearly, the mapping of the previous paragraph is not unitary, but it can be made so 
by supplying an {\it ancilla} prepared in an equal  superposition of all 
natural unit states $\ket{u_l^m},\; u_l^m\{l'\}=\delta_{l,l'},$ 
with $l,l'=1.. m(m-1)/2.$\cite{BAR}

\begin{figure}
\epsfig{file=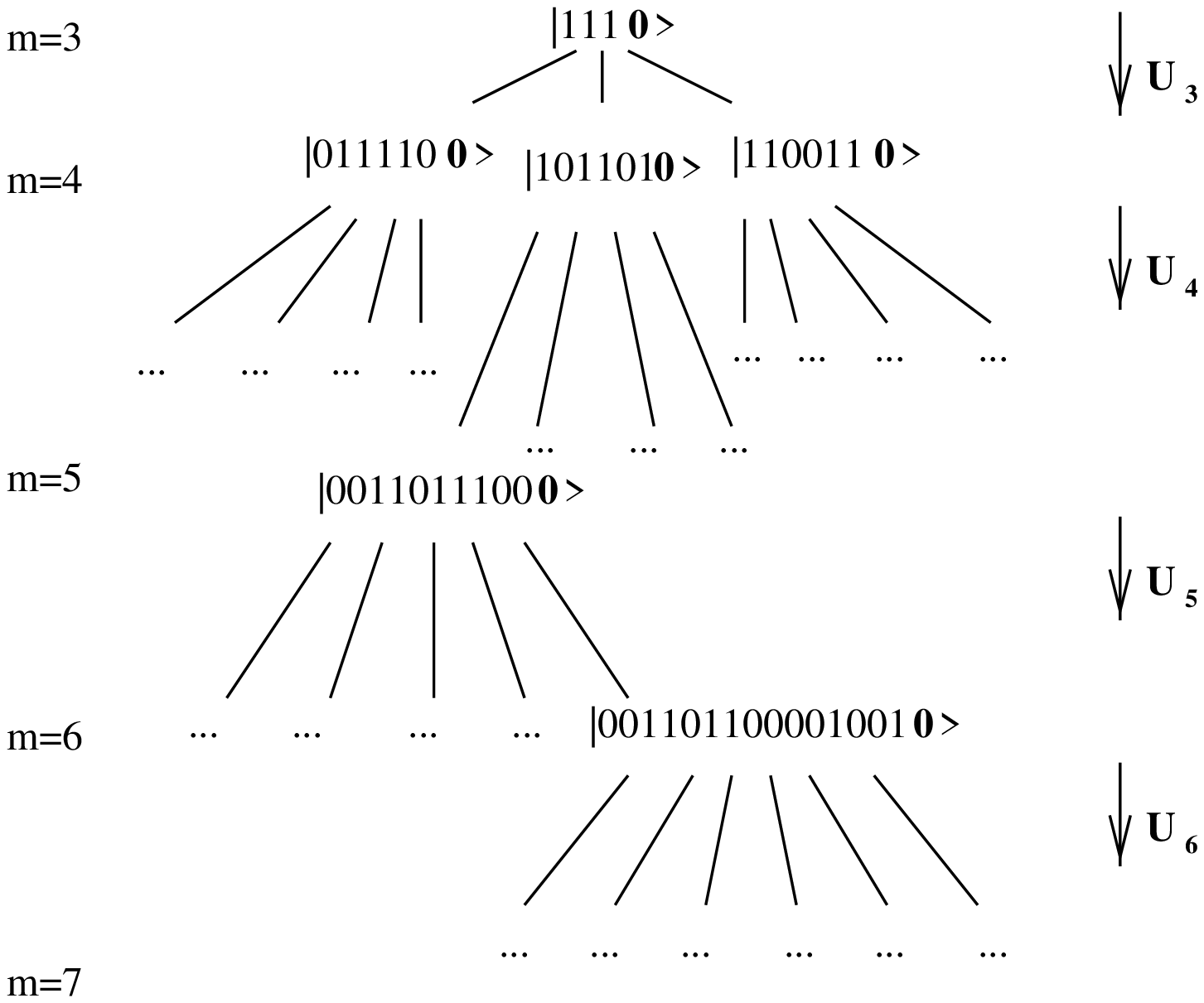,width=6.0in}
~\\
\caption{Partially expanded solution construction tree for TSP,
with Hamiltonian cycles coded by set qbits representing their edges, where
levels of a tree are labeled by the number of vertices $m$.
$U_m$ is a sought mapping between levels.$\bbox{0}$ is
used to denote that all other qbits of a state $\ket{p_k^m}$ are 0.
{\it Ancillae}, necessary at all levels for unitarity of $U_m$ are not
shown.   
}
\label{tree}
\end{figure}

Let us assume for a moment that our system  is prepared in an equal 
superposition of all Hamiltonian cycles over first $m$ vertices:
\begin{equation}
\ket{p}=\frac{1}{\sqrt{(m-1)!/2}} \sum_{k=1}^{m} \ket{p_k^m}.
\end{equation}
Qbit set to 1 in {\it ancilla}  (\mbox{$u_l^m\{l\}=1$})  
points to the edge  in a tour of the input level which should be
broken (if present i.e. if $p_k^m\{l\}=1$) by the insertion of a vertex
of the next level.
On output both these qbits are reset to 0, while setting two bits 
representing newly created edges $p_{k'}^{m+1}\{i\}=p_{k'}^{m+1}\{i'\}=1.$
All other bits of the input path remain unchanged, since the edges they
represent are common for $p_k^m$ and $p_{k'}^{m+1}$.
%It is easy to
%check that we arrive at all $(m)!/2$ Hamiltonian cycles over $m+1$ vertices.
Partially expanded construction tree is depicted in Fig. \ref{tree}, with 
levels labeled by the number of vertices.

Thus defined $U_m$ is an identity  operator except for the  matrix elements:  
\begin{mathletters}
\label{inners}
\begin{eqnarray}
\inner{u_l^m,p_k^m}{\bbox{0},p_{k'}^{m+1}}=
\inner{\bbox{0},p_{k'}^{m+1}}{u_l^m,p_k^m}=1,\\
\inner{\bbox{0},p_{k'}^{m+1}}{\bbox{0},p_{k'}^{m+1}}=
\inner{u_l^m,p_k^m}{u_l^m,p_k^m}=0,
\end{eqnarray}
\end{mathletters}
for such $k.k'$ that:
\begin{eqnarray}
\nonumber
p_k^m\{l\}=u_l^m\{l\}&=\,1,\;p_{k'}^{m+1}\{l\}=0,\\ \nonumber  
p_{k'}^{m+1}\{i\}\,=&p_{k'}^{m+1}\{i'\}=1,
\end{eqnarray}
where, if $l=1.. m(m-1)/2\;$ is a position of a qbit representing the edge $e_{l',l''},$
$i,i'$ are the positions of
qbits representing the edges $e_{m,l''}$, $e_{m,l'}$ and  all other bits of
$p_k^m$ are equal to respective bits of $p_{k'}^{m+1}$ and $\bbox{0}$
denotes that all  the bits of ancilla are 0.

Note that within an adopted convention there is one-to-one correspondence
between indices of a given edge and a position of the relevant bit in a 
list, which can be determined in polynomial time:\newline
In order to find indices $i,k$ of an edge, coded as an $l$-th bit, add the
integers $1+2+\dots$ until the sum is greater than or equal to $l$.
Number of terms in this sum plus one equals $i$. Subtracting the last term
from the sum and subtracting the result from $l$ yields $k$.   
%ADVECTION, data blind, QTCA  

To complete the construction,
we will use the following example:

\paragraph*{Example 1.\label{example}}
We want to construct the unitary transformation, mapping the only cycle
over three vertices: $\ket{111\bbox{0}}$
to the superposition of three possible tours connecting four vertices
using ancilla:
\begin{equation}
   \ket{u_3}=\frac{1}{\sqrt{3}} \left(\ket{100}
                             +\ket{010}
                             +\ket{001}\right)
\end{equation}

and acting with unitary transformation $U_m$ in a product of Hilbert spaces
of our primary set of qbits {${\cal H}_p$} and {\it ancilla} {${\cal H}_u$}:
\begin{eqnarray}
   \frac{1}{\sqrt{3}} (\ket{100}
                             +\ket{010}
                             +\ket{001})\ket{1\;11\;{\bf 0}}
 \stackrel{U_3}{\rightarrow}\frac{1}{\sqrt{3}}\ket{000}\times \nonumber \\ 
\times(\ket{0\;11\;110\;{\bf 0}}+\ket{1\;01\;101\;{\bf 0}}+\ket{1\;10\;001\;
{\bf 0}}) 
\end{eqnarray}

\begin{figure}
\epsfig{file=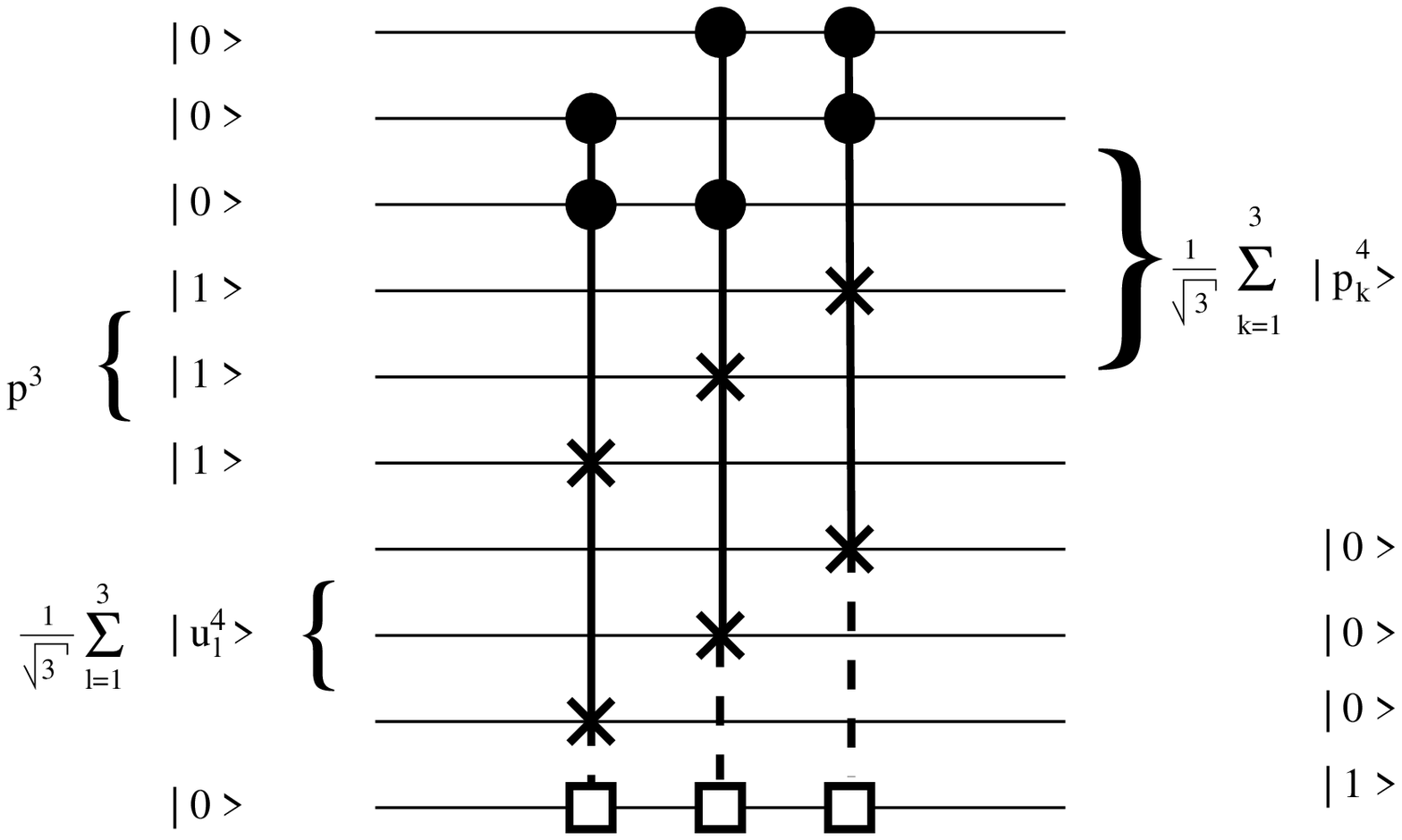,width=6.0in}
~\\
\caption{Factoring of $U_m$ into the bitwise operations in the
case of $m=3$. Full circles and crosses denote respectively change from 0 to 1
and from 1 to 0 according to Eqn.\ref{inners}. For each of thus defined
 4-bit operations affected bits of $\ket{p_k}$ (positions of circles) can be determined from the
set bit of the {\it ancilla} $u_l$ (a position of a cross) in polynomial time. 
Dashed line  and a squares denote a possible extention of $U_m$ and
an associated auxiliary qbit used to faciliate the 
measurements, necessary  at all other levels of the construction.}  
\label{bitwise}
\end{figure}

We can construct this mapping in polynomial time for arbitrary $m$,
factoring
$U_m$ into $m(m-1)/2$ operations $U_m^l$ by iterating over bits of ancilla
and, for each l in turn, performing the four bit operation  $U_m^l$ on
affected qbits of $p$ and a bit of {\it ancilla} as defined by relations 
of Eqs. \ref{inners}.
This construction for the above example is depicted in Fig. \ref{bitwise}.

Thus, in order to recursively implement this construction
to produce the  superposition of all cycles over n vertices we need the time 
${\cal O}(n^3).$ If we do not reuse the {\it ancillae}  the space
requirements will also  be ${\cal O}(n^3)$. 
\paragraph*{Measurement}
Note that the example   describes the only case 
when the valid tour has all bits set, and so the {\it ancilla}
has just right number of states to pass all the amplitude to 
outputs under operation of $U_3$.  

In all subsequent levels some portion of the amplitude will remain in the
input level. We can  correct this by the measurement with projector 
$M_m=\proj{\bbox{ 0 }_u}$ onto state $\ket{\bbox{0}_u}$ of
{\it ancilla} after using $U_m$. Since  {\it all}  
cycles of the output level are entangled with $\ket{\bbox {0}_u}$
the perfect measurement does not destroy the superposition of 
the output tours, while reducing to 0 the amplitude remaining in the 
input level tours.

To establish the scaling behavior of the measurement part of the procedure
let us consider in detail the application of $U_m$ to the state of our system

\begin{eqnarray}
\frac{1}{\sqrt{\frac{m(m-1)}{2}}}\frac{1}{\sqrt{\frac{(m-1)!}{2}}}U_m
\sum_{l=1}^{m(m-1)/2}\sum_{k=1}^{(m-1)!/2}\ket{u_l^m}\ket{p_k^m} = \nonumber \\ 
=\frac{2}{\sqrt{(m-1)m!}}\left( \sum_{\tilde{k}}
\ket{\bbox{0}_u,p_{\tilde{k}}^{m+1}}+\sum_{\bar{l},\bar{k}}\ket{u_{\bar{l}},
p_{\bar{k}}^m}\right) 
\end{eqnarray}
where $\tilde{k}$ is indexing  all the states $\ket{p_{\tilde{k}}^{m+1}}$
fulfilling Eqs. \ref{inners}. They are $m$ qbits set to 1 in each
$\ket{p_k^m}$ 
so  they are \mbox{$m(m-1)!/2= m!/2$} such states --- which can be expected, since
it is the number of distinct cycles  at the $m+1$ level. The indices 
$\bar{k}, \bar{l}$ run through the states where the set bit of {\it ancilla}
corresponds to unset bit of $\ket{p_k^m}.$ 
    
The measurement with $M_m$ reduces the state of our primary
register to:
\begin{equation} 
\ket{p}=\frac{1}{\sqrt{m!/2}}\sum_{k'}\ket{p_{k'}^{m+1}},
\end{equation}
compatible with 
the {\it ancilla} in a one dimensional subspace of ${\cal H}_u$ defined  
by the eigenvector  belonging to the eigenvalue 1 of the projector.
The  probability of finding the system in the desired state is:
\begin{equation}
 P_m=\frac{m}{m(m-1)/2}=\frac{2}{m-1}
\end{equation}

If a measurement of any state of ancilla requires time $T$,
the measurement with $M_m$ will yield the result
in average time $t=T/P_m=(m-1)T/2$.

Thus, total measurement time for the construction of $n$-vertex cycles
is ${\cal O}(n^2)$.
To simplify the measurement we can use an auxiliary qbit $\ket{a_m}$ 
holding the result of a function  $AND(p_m\{l\},u_l^m\{l\})$ (represented by a square in Fig. \ref{bitwise}) calculated
for each $\ket{p^m_k,u_l^m}$. As far as the state reduction is concerned,
the  measurement with $\proj{1_a}$ is equivalent to the  measurement with $M_m.$  

The reversibility of our
construction has special meaning, due to the irreversible measurements $M_m$:
application of $U_m^{\dagger}$ 
to the final state \mbox{$\ket{\bbox{0},p^{m+1}}$} 
results in each $p_k^m$ entangled with all $u_l^m$ representing set qbits of
the cycle, but nevertheless can be used to push the amplitude back to a
single state in $\ket{p}$, while leaving the information of the  cycles in
(entangled) {\it ancillae}.     

The present construction is not time nor space optimal, but it is very 
simple ---
it is not using all the resources postulated for quantum computation
so far and confines much of a calculation to isolated space, thus reducing
errors.
 We need only real amplitudes, and our dynamics is truly digital:
matrix elements of our unitary operators are only 0 or 1. What is more, the
polynomial amplitude errors will lead only to the polynomial 
increase of expected running time, and changing of phases between the
states $\ket{p_k^m}$ of a given level will not  affect the computation at all.

In principle we could have calculated the total weight for each path, but
the probability of finding a minimum in single measurement will be vanishingly
small for large $n.$ 
We could have used {\it ancillae} better matching the requirements of 
unitarity  and, instead of the postulated measurement, the
measurement with a sum of elementary projectors onto  all eigenvectors
of associated operator (repeating $U_m$ with new {\it ancilla} 
n case that the resulting state of ancilla is not $\ket{\bbox{ 0 }_u}$, 
much like in quantum error correction).  

However, the application of the described mapping and postulated
method of state  reduction will be very useful for polynomial 
time algorithms for ${\cal NP}$-complete problems, to 
be presented in a follow-up paper.

The author wishes to thank N. Margolus for guidance into a field of reversible
computation and his hospitality at MIT, where this work originated and to
gratefully acknowledge a valuable discussion with W. Zurek.


\begin{references}
\bibitem{BAR}A procedure for preparing this superposition in time polynomial in $m$
can be deduced from the one described in Eqs.6-8  of \cite{Bar},
here we only note that in an array of spins it can be induced by 
appropriately focused $\pi-$pulse. The state is robust in a sense
that decoherence will require energy exchange with the environment.
\end{references}
\end{document}